\documentclass[sigconf]{aamas}

\usepackage{balance} 

\makeatletter
\gdef\@copyrightpermission{
  \begin{minipage}{0.2\columnwidth}
   \href{https://creativecommons.org/licenses/by/4.0/}{\includegraphics[width=0.90\textwidth]{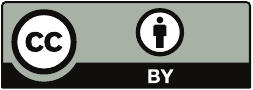}}
  \end{minipage}\hfill
  \begin{minipage}{0.8\columnwidth}
   \href{https://creativecommons.org/licenses/by/4.0/}{This work is licensed under a Creative Commons Attribution International 4.0 License.}
  \end{minipage}
  \vspace{5pt}
}
\makeatother

\setcopyright{ifaamas}
\acmConference[AAMAS '25]{Proc.\@ of the 24th International Conference
on Autonomous Agents and Multiagent Systems (AAMAS 2025)}{May 19 -- 23, 2025}
{Detroit, Michigan, USA}{Y.~Vorobeychik, S.~Das, A.~Nowé  (eds.)}
\copyrightyear{2025}
\acmYear{2025}
\acmDOI{}
\acmPrice{}
\acmISBN{}

\acmSubmissionID{78}

\title[AAMAS-2025 Formatting Instructions]{Multi-Agent Reinforcement Learning Simulation for Environmental Policy Synthesis}
\subtitle{Blue Sky Ideas Track}

\author{James Rudd-Jones}
\affiliation{
  \institution{Centre for Artificial Intelligence, Department of Computer Science, University College London}
  \city{London}
  \country{UK}}
\email{james.rudd-jones.22@ucl.ac.uk}

\author{Mirco Musolesi}
\affiliation{
  \institution{Centre for Artificial Intelligence, Department of Computer Science, University College London}
  \city{London}
  \country{UK}}
\affiliation{
\institution{
Department of Computer Science and Engineering, University of Bologna}
\city{Bologna}
\country{Italy}}
\email{m.musolesi@ucl.ac.uk}

\author{María Pérez-Ortiz}
\affiliation{
  \institution{Centre for Artificial Intelligence, Department of Computer Science, University College London}
  \city{London}
  \country{UK}}
\email{maria.perez@ucl.ac.uk}

\begin{abstract}
Climate policy development faces significant challenges due to deep uncertainty, complex system dynamics, and competing stakeholder interests. Climate simulation methods, such as Earth System Models, have become valuable tools for policy exploration. However, their typical use is for evaluating potential polices, rather than directly synthesizing them. The problem can be inverted to optimize for policy pathways, but the traditional optimization approaches often struggle with non-linear dynamics, heterogeneous agents, and comprehensive uncertainty quantification. We propose a framework for augmenting climate simulations with Multi-Agent Reinforcement Learning (MARL) to address these limitations. 
We identify key challenges at the interface between climate simulations and the application of MARL in the context of policy synthesis, including reward definition, scalability with increasing agents and state spaces, uncertainty propagation across linked systems, and solution validation. Additionally, we discuss challenges in making MARL-derived solutions interpretable and useful for policy-makers. Our framework provides a foundation for more sophisticated climate policy exploration while acknowledging important limitations and areas for future research.

\end{abstract}


\newcommand{\BibTeX}{\rm B\kern-.05em{\sc i\kern-.025em b}\kern-.08em\TeX}

\begin{document}


\pagestyle{fancy}
\fancyhead{}


\maketitle

\section{Introduction} 
Climate policy derivation represents one of society's most difficult governance challenges, characterized by deep uncertainty, competing stakeholder interests, and complex interdependencies across social, economic, and environmental systems \citep{stern2022economics}. 
At the national level, climate policy formulation typically involves analyzing emissions reduction potential, assessment of technological pathways, and evaluation of policy actions \citep{hulme2004understanding}.
Multi-stakeholders are accounted for to ensure a cohesive approach, but extending to the continental or global scale makes this process more challenging.
This conventional approach faces further fundamental difficulties due to the socio-environmental domain that can be summarized threefold.
Firstly, the global climate system has significant temporal latency between action and downstream effects.
Many of these effects are not monitored or hard to capture, further increasing this latency and making causal links between policy implementation and observable outcomes hard to solidify \citep{matthews2020opportunities}. 
This temporal disconnect can span decades, making it very challenging to evaluate policy effectiveness and adjust strategies in response to emerging data \citep{ricke2014maximum}. 
Secondly, climate policies often generate unevenly distributed impacts across different socioeconomic groups and regions, potentially exacerbating existing inequalities and creating political resistance \citep{diffenbaugh2019global}. 
Thirdly, the presence of tipping points and feedback loops in the Earth system introduces non-linear dynamics that traditional policy analysis struggles to address \citep{lenton2019climate}.
Current approaches to climate policy development rely heavily on international negotiations and evidence synthesis from various scientific disciplines. Organizations such as the IPCC coordinate massive efforts to provide policy-relevant insights \citep{ipcc}. However, this process can be slow, politically constrained, and sometimes fails to capture the full range of possible policy interventions \citep{victor2019accelerating}, including the deep uncertainty inherent in climate projections \citep{nordhaus2013climate}.

\begin{figure*}[h]
  \centering
  \includegraphics[width=1.0\linewidth]{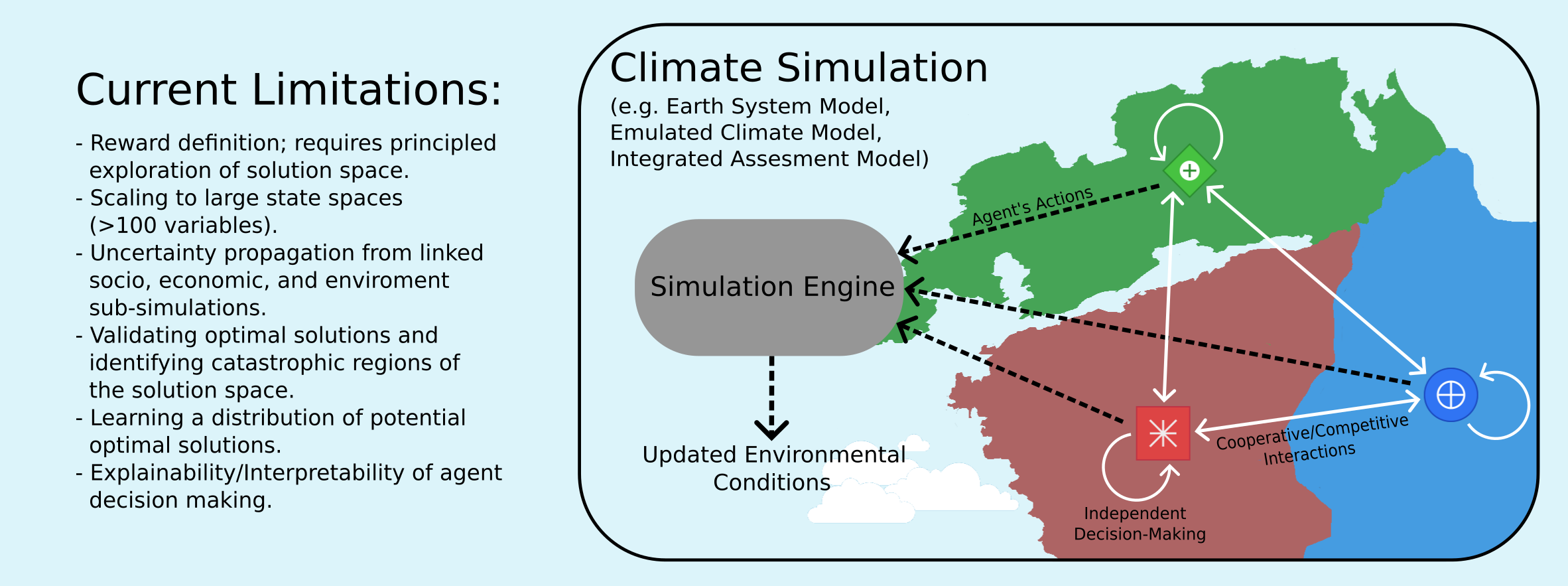}
  \caption{On the right: A conceptual framework of a multi-agent climate simulation system. The outer climate simulation is imbued with socio-economic agents, in this case three geographical regions. Agents make independent decisions while engaging in cooperative/competitive interactions (white arrows). A central simulation engine (grey) processes agents' actions (dashed black lines) for the linked socio, economic, and environmental simulations, updating the environmental conditions accordingly. On the left: An enumeration of six key technical open challenges currently facing such systems.}
  \label{fig:rect1}
  \Description{Logo of AAMAS 2025 -- The 24th International Conference on Autonomous Agents and Multiagent Systems.}
\end{figure*}

Simulation-based approaches have emerged as crucial tools for addressing these limitations. They enable policy-makers to explore potential outcomes in a risk-free environment, facilitating the evaluation of various policy combinations and their long-term implications \citep{van2013cross}. 
Being able to quantify the forecasted effects of policies not only guides the policy derivation process but also provides evidence to address concerns raised by critics \citep{cairney2016politics}. 
Earth System Models (ESMs) provide a crucial simulation framework, offering high-resolution representations of atmospheric, oceanic, and terrestrial processes \citep{flato2014evaluation}. 
For example, they have directly informed national carbon budgets and adaptation strategies by simulating changes in temperature, precipitation, and extreme weather events \citep{shepherd2018storylines}. 
However, their utility in policy-making is generally constrained as the influence of society is often not an internal process in these models.
Exogenous variables, such as atmospheric carbon, are changed over time to ``simulate'' anthropogenic effects, with limited feedback on how the changing climate would affect society \citep{steffen2018trajectories}.
Instead, Integrated Assessment Models (IAMs) combine socio-economic and environmental components to model climate socio-economic interactions, representing the current state-of-the-art in climate policy simulation. 
Generally, they use simplified environmental processes derived from ESMs but the incorporation of linked socio-economic simulations allows for a more defined feedback loop between the environment and society.
Notable examples include the DICE model \citep{nordhaus2017revisiting}, REMIND \citep{luderer2013remind}, and GEMINI-E3 \citep{bernard2008gemini}. 
On the UN website they publicly list twenty-nine IAMs used for their decision making \citep{Integrat69:online}, as an example GEMINI-E3 was used to analyze the future $CO_2$ emission trajectories of the national policies made at the COP26 Glasgow conference \cite{van2023multimodel}.
IAMS can be preferable to ESMs as it allows a more direct way to evaluate policy interventions as cascading effects between environment and society are linked.
Importantly the problem can be ``flipped''; \textit{we can optimize for an idealistic policy}.
We focus in on IAMs due to this feature, as in ESMs this is not as connected. In ESMs we could define optimizing criteria, but the resultant solution has no explicit connection to a policy and there are no clear ``actions''.
Traditional optimization approaches for IAMs (for example Model Predictive Control \citep{garcia1989MPC}) solve for a set utility function.  
However, may rely on simplifying assumptions, discrete state spaces, and local linear approximations to make the problem computationally tractable \citep{deffuant2011viability, strnad2019deep}. 
These solvers typically only explore a limited portion of the possible solution space, potentially missing innovative policy combinations \citep{otto2020social}.
On top of this, most models use highly aggregated representations of socio-economic agents, failing to capture the diversity of stakeholder behaviors and interactions \citep{mercure2016modelling}.
Finally, these approaches in IAMs often focus on finding a single ``optimal'' solution, which may not adequately capture the full range of possible outcomes \citep{weyant2017some}. This in itself can fail to capture some notion of uncertainty that exists in both climate system responses and socioeconomic developments.

These limitations point to the need for more sophisticated approaches that can better handle uncertainty, complex system dynamics, interactions between heterogeneous agents, and the vast solution space of possible policy combinations. Modern computational techniques, particularly Reinforcement Learning (RL) and Multi-Agent Reinforcement Learning (MARL), offer promising capabilities to address these challenges. 
RL approaches can overcome these limitations by providing more robust exploration of policy spaces, not requiring linear assumptions, and improved handling of uncertainty. MARL approaches can better represent agent heterogeneity for resilient policies. 
For a single agent scenario, the recent work of \cite{strnad2019deep} and \cite{wolf2022climate} applied an RL agent into an IAM, that generated policy guidance pathways towards a defined ``economic and environmental positive future''. Both \cite{strnad2019deep} and \cite{wolf2022climate} use a singular agent, hence assuming a ``unified'' earth, in which there is a collectively shared goal, restricting the potential for agent heterogeneity.
For agent heterogeneity, \cite{zhang2022ai} created the RICE-N model, itself an extension of the Regional Integrated model of Climate and the Economy (RICE) model developed in \cite{nordhaus2010economic} that models twelve global regions. The RICE-N model imbues an IAM with trade and negotiation dynamics for further interactions between agents \citep{zhang2022ai}. 
Further, \cite{ruddjones2024ayseds} extended the environment in \cite{wolf2022climate} towards multiple agents, using the intepretability gained from the low-dimensional environment to understand policy pathways. 

\section{Towards a Framework for MARL derived Climate Policy}  
ESMs and IAMs can be used to validate potential policy interactions in a safe simulated space. More importantly we can work the other way and derive optimal policies.
Whilst ESMs can be anthropogenically influenced, we focus on IAMs as a case study since they have a more direct connection between socio and environmental systems that better suite optimization. 
RL is able to model non-linear stochastic systems, greatly aligning its use with IAM dynamical systems, and accounting for long term reward well posits its use in policy trajectory derivation.
Importantly as a first step, how can we reformulate IAMs as an RL problem, or more concretely can we setup an IAM as a Markov Decision Process (MDP) \citep{puterman2014markov} for RL or Stochastic Game (SG) \citep{shapley1953stochastic} for MARL?
IAMs already share significant structural similarities with RL environments; they inherently require objective functions if optimized, that can be used as a reward function.
Further, actions can be defined as varying model parameters over time to reflect some policy trajectory.
For example, investment decisions, policy implementations, and technological adoption rates.
To summarize, all that is needed to utilize RL/MARL with an IAM is defining a set of actions, and accessing outputs of a running simulator or porting the simulator into a programming language that is suited for RL/MARL research.

\noindent \textbf{Why Use MARL?} 
Anthropogenic climate change inherently involves heterogeneous actors with mixed motives, strategic interactions, and complex cooperation-competition dynamics - elements that single-agent RL cannot adequately capture. 
Mirroring major criticisms in the IAM literature where economic and behavioral dynamics are poorly represented by a singular entity \citep{mercure2016modelling, madani2013modeling}. 
Further, MARL's ability to model long-term strategic behavior while accounting for multiple interacting entities makes it particularly suitable for climate policy modeling \citep{leibo2017multi, hughes2018inequity}.

\noindent\textbf{What are we Optimizing with MARL?} A clarification upon moving to MARL is how we view the optimization. We believe there are two main views that necessitate certain restrictions on the MARL application.
\textit{Firstly}, one could focus on purely modeling all heterogeneous agents interacting together - we want to understand the dynamics between agents as some form of sequential social dilemma (SSD). 
MARL has emerged as a powerful framework for studying SSDs and ``tragedy of the commons'' type scenarios, offering insights into emergent social dynamics and equilibrium behavior. Unlike traditional game-theoretic approaches that often rely on one-shot interactions, MARL enables repeated sequential decisions with complex state dynamics \citep{leibo2017multi}. This approach has proven particularly valuable in commons problems, where agents must balance immediate individual rewards against long-term collective welfare \citep{hughes2018inequity}. 
These studies collectively suggest that MARL not only serves as a tool for finding Game-Theoretic equilibria without heavy assumptions but also provides insights into the mechanisms that drive the evolution of social behavior and cooperation in complex, dynamic environments \citep{peysakhovich2018consequentialist, tacchetti2019relational}.
Results from this view-point can help domain specialists understand the effects of certain simulation or agent parameters on an IAM equilibrium. 

\textit{Secondly}, a more applicative approach entails planning policy pathways for one or a subset of the agents, and using representative models of other entities. 
The resultant pathways can be used to guide decisions by policy makers, as the framework provides evidence of their validity.
As an example use case, the other agents in the simulation (that we have no agency over in reality) could be trained using imitation learning on historical data to represent in-silico versions of real-world entities.
The MARL agent(s) can learn a best response to these pre-trained agents dependent on the validity not only of the IAM, but also the representative agents.
This approach shares similarities with the domain of Ad-Hoc Teamwork \citep{stone2010ad, wang2024n_player_ad_hoc, wang2024open}, which explores how agents can learn to interact with novel and unknown partners.
This second framework necessitates the use of decentralized algorithm approaches as agents must not share model parameters, rewards, or observations.

\section{Open Challenges}

\noindent\textbf{Reward Definition.} A common issue with any optimization approach is the objective function, which requires some bias towards a chosen target variable or parameter \citep{pindyck2017use}. Motivating these is tricky and is up to the modelers discretion \citep{pindyck2017use}.
Similarly for RL and MARL we have the definition of a reward function, which can be an arbitrary choice. 
There are two sides of the coin here, rewards in RL/MARL can be more abstract than objective functions. For example a binary reward for reaching a certain end state, that encodes complex objectives like sustainability and equity, but may mask important nuances of the underlying simulation \citep{silver2021reward}.
We could set rich reward signals such as directly relating to a normalized economic or environmental variable.
In this domain even dense rewards may receive many sequential negative ``poor'' rewards (e.g., strict legislation) until a ``positive'' outcome is reached.
This is a typical deep exploration task in RL (e.g. deepsea \citep{osband2019behaviour}), which generally require uncertainty based exploration methods to fully search the solution space \citep{fellows2021bayesian_thesis}.
Uncertainty based exploration in RL is an extensive field \citep{osband2018randomized, o2023efficient, fellows2023bayesianexplorationnetwrosks}, but is not so explored in MARL \citep{hao2023exploration, mahajan2019maven, zintgraf2021meliba, schafer2023ensemble}.
Due to the non-stationarity in the transition function there seems even more of a need for principled exploration. 
On top of this, the climate can change rapidly as tipping points are reached requiring quick behavioral responses from the other agents, both exacerbating non-stationarity \citep{farahbakhsh2024tipping}.

\noindent \textbf{Scalability.} RL and MARL approaches can be more adept in the complex IAM environment over traditional optimal control approaches, but can similarly struggle with scalability.
Particularly in MARL where increasing the number of agents can lead to an exponential growth of state and action spaces in centralized training approaches \citep{yu2022surprising}.
Decentralized approaches which are required for the \textit{second MARL viewpoint}, are generally able to scale more easily as model parameters are not shared. 
However, coordination becomes more challenging and thus requires more complex methodologies that impact scaling, especially as numbers of agents grow beyond hundreds.
Further compounding the issue, large scale IAMs such as the Intertemporal General Equilibrium Model \citep{nordhaus2013integrated} that has 4000 endogenous variables, lead to a very large state space for any RL application.
Reaching these IAMs not only raises challenges in the algorithmic side, but these large-scale simulators require much larger computational budgets, further increasing training times.
The difficulty lies in coupling these systems - economic shocks influence social responses, which affect climate policy implementation, creating complex cascade effects \citep{ming2023deep}.

\noindent \textbf{Uncertainty Representation.} Understanding model uncertainty is inherent to the downstream use, it is imperative to understand when our temporal predictions become no better than random choices. 
IAMs contend with significant sources of uncertainty from many sources.
Each separate socio-economic and environmental model has its own epistemic uncertainty stemming from modeling assumptions and aleatoric uncertainty from inherent noise in the underlying process.
As these models are all linked there is further uncertainty in their interactions, particularly climate sensitivity, damage functions, and economic growth projections \citep{pindyck2013, gillingham2015}. 
Further, IAMs bridge multiple temporal and spatial scales, from short-term economic decisions to long-term climate dynamics. Uncertainty compounds across these scales, making it difficult to accurately represent interactions between local impacts and global processes \citep{vanderploeg2019}.
Decision making becomes intractable as many climate risks exhibit heavy-tailed probability distributions \cite{weitzman2011}. 
As discussed earlier, uncertainty-driven methods in RL can bring deeper exploration of the solution space and have a chance to adapt to these regions of greater uncertainty. 
However, typically exploration is guided by epistemic uncertainty of the RL agent and does not account for the aleatoric uncertainty stemming from the simulator \citep{fellows2023bayesianexplorationnetwrosks}. 
Importantly, how can we factor in the multiple sources of uncertainty from the linked models, to get better calibrated notions of uncertainty in the suggested solutions?

\noindent\textbf{Solution Validation.} Validating adaptions of IAMs to MDPs or SGs can be fairly straightforward, simulation dynamics can be compared, usually with minimal discrepancy as shown in \cite{ruddjones2024ayseds}. Comparing traditional optimization approaches with the RL and MARL solution is fairly similar. 
Distance metrics between potential pathways, or final visitation states quantify the performance of the framework. 
The real challenge lies in validating simulation solutions with real world applicability, connecting to broader questions in simulation intelligence and model validation \citep{lavin2021simulation}. 
Perfect validation against real-world outcomes, particularly for long-term climate predictions, is fraught with risk due to the inherent stochasticity of such systems.
Importantly, it can be easier to identify and validate the unfeasible or undesirable trajectories, which may enter areas of the solution space that we can have more assurance on being incorrect.
Mapping out dangerous trajectories provides valuable insights for decision-makers to avoid catastrophic scenarios. 
Quantifying and understanding negative outcomes can be as crucial as identifying optimal solutions.

\noindent \textbf{Distribution of Solutions.} Traditional optimization approaches usually provide only one optimal solution without describing the robustness of said approach \citep{weyant2017some}.
Due to the inherent uncertainties of IAMs, and the challenges from exactly following a projected pathway, solutions that have more tolerance for error, in that they are more resilient to shocks or tipping points, are preferable.
In RL/MARL this is possible by evaluating optimal agent policies with various initial states to get a distribution of solutions.
It can be a cumbersome approach, as if the optimal solution is found to be non-robust, the agents must be retrained and the process iterated.
Instead recent work by \cite{leondiscovering} finds a diverse range of optimal solutions in complex highly dynamic environments. 

\noindent \textbf{Explainability \& Intepretability} - 
Explainability is an extensive challenge in any use of Machine Learning, but especially so in RL/MARL when using complex non-linear models. While traditional control methods like MPC offer clearer insights into their decision-making through explicit optimization objectives and constraints, deep RL methods often operate as black boxes.  
Most approaches provide post-hoc explanations that attempt to provide understanding between agent actions and certain states.
Our framework provides policy trajectories in the solution space of the IAM, but these larger state spaces easily extend beyond human interpretable three-dimensional spaces. 
How are downstream users able to interpret solutions, are we able to visualize policy trajectories in high-dimensional spaces?

\section{Limitations}

Primarily, the MARL solution can only be as valid as the underlying simulator; improved optimization solutions cannot transcend the constraints of an imperfect model.
This becomes particularly problematic in these environmental domains characterized by unprecedented scenarios or complex emergent phenomena, such as extreme weather patterns as we pass climate tipping points \citep{farahbakhsh2024tipping}. 
Current climate simulators may struggle to accurately model the feedback loops between anthropogenic caused rising temperatures and the downstream effects, leading to potentially dangerous blind spots in MARL-based policy recommendations.
While there has been progress in open-ended learning within RL frameworks \citep{wang2020enhanced}, this capability must be matched by equally adaptable simulation environments - a requirement that may exceed current modeling capabilities. 
However, these limitations do not entirely diminish the value of our approach; since our framework is agnostic to the underlying world system model (ESM or IAM or more), it can readily incorporate improved simulators as they become available, ensuring adaptability across various domains and modeling paradigms.


\section{Conclusion}


This paper presents a framework for enhancing climate policy exploration through the integration of MARL with IAMs. While MARL offers promising capabilities for handling non-linear dynamics, agent heterogeneity, and uncertainty quantification, significant challenges remain. The framework's success hinges on addressing key challenges including principled exploration in sparse reward settings, scalability of both algorithms and simulations, and the propagation of uncertainty across coupled systems. Moreover, translating MARL-derived solutions into actionable policies requires advances in explainability and visualization techniques. We highlight three promising research directions: \textbf{1.} uncertainty-driven MARL algorithms to handle the multiple sources of uncertainty inherent in IAMs; \textbf{2.} MARL for extensive state spaces that may rely upon linked emulations; \textbf{3.} explainable RL techniques specifically tailored for large state spaces.
We hope that this position paper drives efforts towards the improvement of simulation derived climate policy, supporting political guidance to return the Earth's trajectory onto a habitable and stable future.

\begin{acks}
James Rudd-Jones is supported by grants from the UK EPSRC-DTP (Award 2868483). 
Mirco Musolesi was supported by the Italian Ministry of University and Research (MUR) through the project PRIN 2022 “Machine-learning based control of complex multi-agent systems for search and rescue operations in natural disasters (MENTOR)” funded by the European Union - NextGenerationEU.
\end{acks}



\bibliographystyle{ACM-Reference-Format} 
\bibliography{sample}


\end{document}